\documentclass[useAMS,usenatbib]{mnras}
\usepackage{graphicx}
\usepackage{subfig}
\usepackage{color}
\usepackage[normalem]{ulem}

\newcommand{\NeX}{Ne \MakeUppercase{{\romannumeral 10}}~}
\newcommand{\NeTen}{Ne$^{10+}$}

\title[\NeX X-ray Emission due to Charge Exchange]{\NeX X-ray Emission due to Charge Exchange in M82}
\author[R. S.\ Cumbee et al.]{R. S.\ Cumbee$^{1}$\thanks{rcumbee@physast.uga.edu},
L. Liu$^{2}$, D. Lyons$^{1}$, D. R. Schultz$^{3}$, P. C. Stancil$^{1}$,  
\newauthor J. G. Wang$^{2}$, and R. Ali$^{4}$\\ 
$^{1}$Department of Physics and Astronomy and the Center for Simulational Physics, University of Georgia, Athens, GA 30605, USA\\
$^{2}$Institute of Applied Physics and Computational Mathematics,  PO Box 8009, Beijing 100088, Peoples Republic of China\\
$^{3}$Department of Physics, University of North Texas, Denton, TX, 76203, USA \\
$^{4}$Department of Physics, The University of Jordan, Amman 11942, Jordan}

\begin{document}

\date{}

\pagerange{\pageref{firstpage}--\pageref{lastpage}} \pubyear{2016}

\maketitle

\label{firstpage}

\begin{abstract}

Recent X-ray observations of star-forming galaxies such as M82 have shown the
Ly$\beta$/Ly$\alpha$ line ratio of Ne X to be in excess of predictions for thermal electron
impact excitation.  Here we demonstrate that the observed line ratio may be due to
charge exchange and can be used to constrain the ion kinetic energy to be  $\la$ 500 eV/u. This is
accomplished by computing spectra and line ratios via a range of theoretical methods
and comparing these to experiments with He over astrophysically relevant collision
energies.  The charge exchange emission spectra calculations were performed for
\NeTen + H and \NeTen + He using widely applied approaches including the
atomic orbital close coupling, classical trajectory Monte Carlo, and multichannel
Landau-Zener (MCLZ) methods.  A comparison of the results from these methods indicates
that for the considered energy range and neutrals (H, He) the so-called ``low-energy
$\ell$-distribution'' MCLZ method provides the most likely reliable predictions.

\end{abstract}


\section{Introduction}

X-ray emission from various solar system environments \citep{cra00, lis96,den10,kra04}
such as comets and planetary atmospheres \citep{den02} has been attributed to de-excitation
of highly excited product states following the process of charge exchange (CX).
Observations have suggested that CX could be a significant contributor to X-ray
emission outside the heliosphere, such as in 
supernova remnants \citep{kat11,cum14}, starburst galaxies \citep{liu11}, and extragalactic 
cooling flows \citep{fab11,lal04}.

Within an environment in which both ions and neutral atoms or molecules are present, CX can occur
and may dominate the ion emission spectra. During a collision, the ion $X^{q+}$ captures
an electron from a neutral atom 
creating a highly excited ion $X^{(q-1)+}(n\ell~^{2S+1}L)$ that then emits one or more
photons in a cascade down to the ground state.
To accurately model the X-ray emission spectrum and hence better understand neutral and ion density distributions, ion temperatures,
and ion charge state distributions in the environments considered,
it is essential to include the dominant collisional processes, including CX, electronic recombination and excitation, and photonic excitation
for different relevant ion and neutral interactions. 
  
For example, X-ray line emission has been shown to be prominent in observations of hot plasma environments, and
\citet{lal04} and \citet{kon11} have suggested that line emission in M82 may partially be due to CX
in which a highly ionized plasma collides with cold gas \citep[see also][]{zha14}. Specifically, \citet{kon11} suggest that CX emission from
\NeX may make a significant contribution to the spectra. A better understanding of the CX emission of \NeX in this environment
could help constrain ion and neutral densities in M82 as well as give insight into what other environments CX is prominent.  

Here we compare CX X-ray emission due to Ne$^{10+}$ collisions with H and He using a number of theoretical methods. In the case of He, X-ray
spectra are compared to a previous experiment that used a coincident X-ray - ion charge state and momentum detection approach 
that allowed determination of the charge exchange product state. Computed Lyman series line ratios are 
compared to the \NeX Ly$\beta$/Ly$\alpha$
ratio in the disk of M82 and are found to be consistent with a CX interpretation.

\section[]{Results and Discussion}

\subsection{Charge-Exchange Cross Sections}

\begin{figure*}
\includegraphics[scale = 0.6, angle = 270]{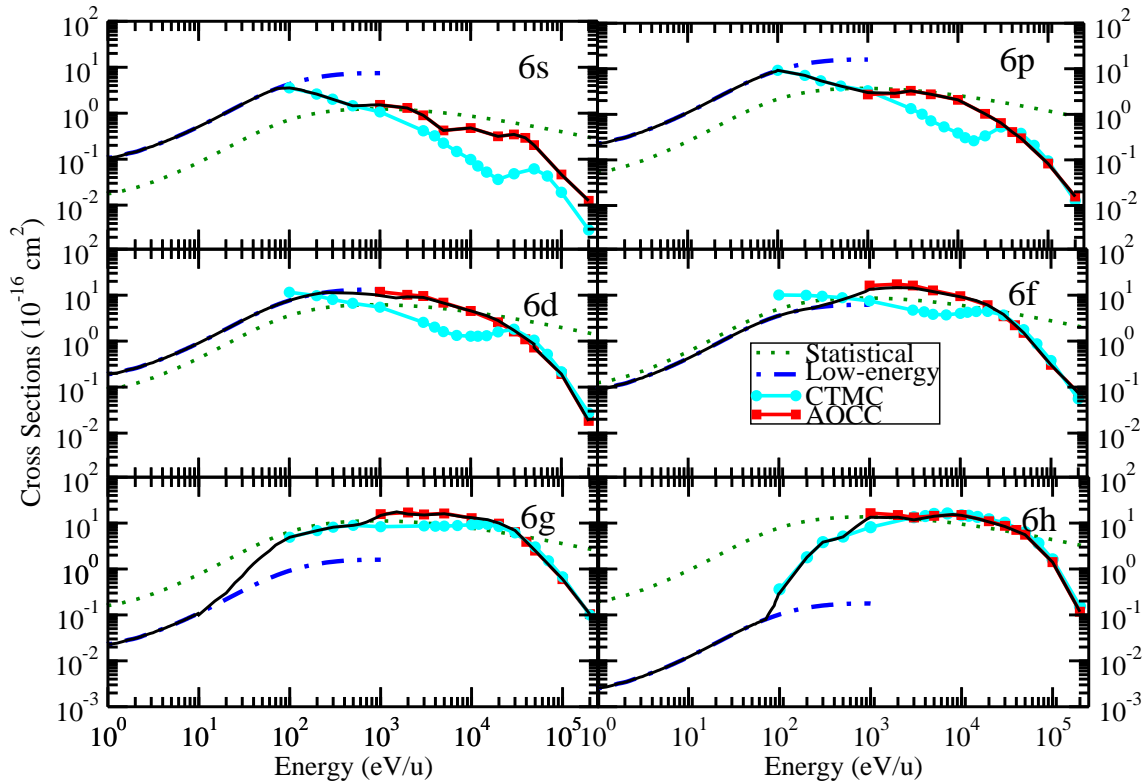}
  \vspace*{1pt}
  \caption{New charge exchange cross sections for the collision of Ne$^{10+}$ with H as a function of collisional energy
  for the dominant capture channel of $n$=6. 
  The blue (-..) lines represent the MCLZ low energy distribution, the green dotted (...) lines represent the MCLZ statistical distribution, the cyan
  lines with filled circles represent the CTMC method, and the red lines with squares represent the AOCC method. The solid black line shows 
  our suggested cross sections.}
\label{FigCX}
\end{figure*}

Cross sections for Ne$^{10+}$ with H and He for capture to state-resolved Ne$^{9+}$($n\ell~^2L$) levels have been calculated using the
classical trajectory Monte Carlo (CTMC), atomic-orbital close-coupling (AOCC), and multi-channel Landau-Zener (MCLZ) methods.
While both collision systems have been studied for many decades \citep[e.g.,][]{sch95,liu14}, none of the previous studies have presented a comprehensive set of 
state-resolved cross sections over a range of energies relevant to astrophysics. 
The \NeTen + H CX cross sections discussed in this paper are shown in Figure~\ref{FigCX} for the dominant capture channel $n=6$. 
Details of the current MCLZ cross section calculations used here are summarized in \citet{lyo14}.

The AOCC method for the description of ion-atom collision processes is discussed in detail 
elsewhere \citep{fri91}. We shall only outline it briefly here. For a one-electron – two-center collision system the total electron wave function 
is expanded over the traveling atomic orbitals centered on each of the two centers (nuclei of the projectile ion and of the target ion). 
Here, the determination of electronic states centered on the
target and on the projectile, is performed by using the variational method with even-tempered trial functions  \citep{kua97,ree63}. 
The total electron wave function  $\Psi$ can be expanded in terms of atomic orbitals, each multiplied by a plane wave electron translational
factor (thus giving a traveling atomic orbital $\Psi$). Inserting $\Psi$ into the time-dependent Schr\"{o}dinger equation of the collision system, 
one can get the first-order coupled equations for the amplitudes centered on each of the nuclei. After solving the coupled equations, one obtains 
the transition amplitudes, whose squared moduli at infinitely large times give the transition probabilities  and from them the state-selective CX cross sections
that were computed for kinetic energies between 1 and 200 keV/u.

The CTMC method solves an $n$-body problem (i.e., consisting of a
projectile ion, target nucleus and inactive target electrons, and active target
electrons) by solving the classical Hamilton's equations for a large ensemble of
projectile-target configurations (``trajectories'') to simulate ion-atom
collisions \citep{abr66,ols77}.

The trajectories
sample initial electronic orbitals prepared to emulate the quantum
mechanical electronic momentum distribution. At the end of the trajectory,
at some asymptotic distance, relative classical binding energies between
particles are computed and used to ascertain whether a reaction (charge
exchange, excitation, or ionization) occurred. In the case of charge
exchange to produce Ne$^{9+}$, the final $n\ell$ quantum state is
determined based on the binding energy and classical angular momentum
of the electron, if bound to the projectile ion, following classical-quantum binning rules \citep{bec84}. 
Cross sections for H collisions were obtained between 0.1 and 200 keV/u.

\subsection{X-ray Spectra}
In order to predict CX-induced spectra,
 a low density, steady-state radiative cascade model was applied as described by \citet{rig02}. In this model, the
initial state populations are proportional to the quantum-state-resolved CX cross sections. As the 
electron cascades down to the lowest energy level, obeying quantum mechanical selection rules, 
photons are emitted, including X-rays.

When the initial charge $q$ of the ion $X^{q+}$ is greater than $\sim$3,
the electron will be primarily captured into an excited state that decays
to a lower state, such as the ground state, and emits at least one photon. 
In the case of Ne$^{10+}$+H that produces the hydrogen-like ion Ne$^{9+}$ primarily
in the principal quantum level $n=6$ \citep[see, for example,][]{kra04, smi14},
the electron can be captured to any orbital angular momentum state from $\ell=0$ ($6s$) to
$\ell=5$ ($6h$) with a range of probabilities. In the case
that the electron is captured to the $6p$ state of Ne$^{9+}$, it will primarily decay to the
$1s$ ground state emitting a 1322 eV X-ray photon, the Ly${\epsilon}$ line.
In another case in which the electron is captured to the $6h$ state,
it can only decay by a series of $\Delta \ell = \Delta n = -1$ Yrast transitions that
result in infrared and visible photons. When the electron arrives at the $2p$ level, it will
decay to the ground state emitting the 1020 eV Ly${\alpha}$ line.
Ly${\beta}$, Ly${\gamma}$, and Ly${\delta}$ lines will also be produced from the cascade
resulting from capture to other 6$\ell$ states.
While $n=6$ is the dominant
manifold for electron capture with H, there are also significant probabilities for
capture to $n=7$ and $n=5$ followed by cascade and
X-ray emission. Therefore, to fully predict the X-ray spectrum of Ne$^{9+}$ following 
CX collisions with H, or similarly with He, requires knowledge of many $n\ell$-resolved
cross sections.

While the basic CX process has been studied for decades,
the existing data are lacking in uniform reliability, in coverage
of projectile/target combinations, and in energy ranges relevant to astrophysics. 
Measurements
of state-selective cross sections (i.e., resolved by $n$ and/or $\ell$) are
rare and difficult to perform.
The energy resolution is typically
poor or the studies are restricted to the measurements of a few
final states due to limited detector wavelength ranges. Improvements have
been made with experiments to directly measure the X-ray spectrum
\citep{gre01}, but the measurement is of the total spectrum including the
photons from all routes of the cascade. 
Hence, to directly determine state-selective cross section data, one
must turn to theoretical methods. 
For product ions with more than one electron, or
for collisions on He, the status and reliability of calculated cross
sections is generally lower than for H. Therefore, there is a need for measured cross sections
and for an estimation of them via empirical methods in addition to theoretically predicted results.

In this paper, spectra were obtained for Ne$^{10+}$ CX collisions with H and He for a range of kinetic energies. 
For Ne$^{10+}$+H CX, cross sections from the AOCC, CTMC,
and MCLZ methods are adopted. For the MCLZ method, calculations were performed using the low energy and statistical 
$\ell$-distributions described in equations~(15) and (16) of  \citet{jan83} \citep[see also][]{kra04, smi14}. 
It is necessary to utilize $\ell$-distribution functions with MCLZ calculations for product H-like ions (i.e., bare projectiles). 
Due to degeneracy of a given $n$-level in a non-relativistic electronic description,
only direct $n$-resolved MCLZ computations are possible \citep{abr77}. 
Calculations for Ne$^{10+}$ + He CX were
performed using only the MCLZ method here, with both the low energy and statistical $\ell$-distributions as with H. 
Nevertheless, the present (and previously published) AOCC and CTMC data
for \NeTen + H CX, applicable primarily at higher impact energies than of
astrophysical relevance in nebular environments, are instructive for comparison with
MCLZ calculations in order to assess their reasonableness in the range of energies
where they have overlapping applicability with CTMC and AOCC.  We note below the
existence of CTMC data for \NeTen + He CX presented previously in \citet{ali10}.

\begin{figure}
\begin{center}
\includegraphics[scale=0.35, angle=270]{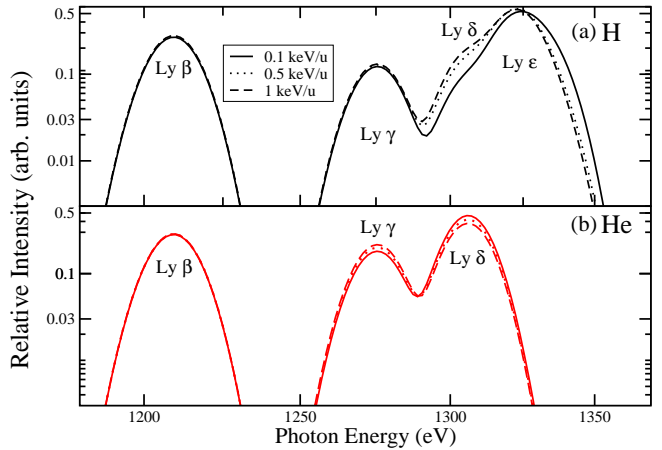}
\end{center}
\caption{ X-ray spectra using the  MCLZ cross sections with the low energy $\ell$-distribution are shown 
for the collisions of Ne$^{10+}$ with H (top panel) and He (bottom panel) at collisional energies of 0.1, 0.5  and 1 keV/u. Lines are normalized 
to Ly$\alpha$ (not shown) at 1020 eV with a resolution of 10 eV FWHM.}
\label{Fig1}
\end{figure}

\begin{figure}
\begin{center}
\includegraphics[scale=0.35, angle=270]{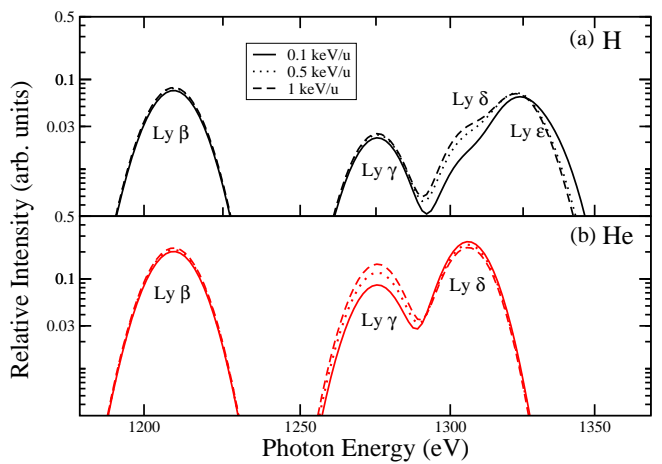}
\end{center}
\caption{ X-ray spectra using the  MCLZ cross sections with the statistical $\ell$-distribution are shown 
for the collisions of Ne$^{10+}$ with H (top panel) and He (bottom panel) at collisional energies of 0.1, 0.5  and 1 keV/u.}
\label{Fig2}
\end{figure}

\begin{figure}
\begin{center}
\includegraphics[scale=0.35, angle=270]{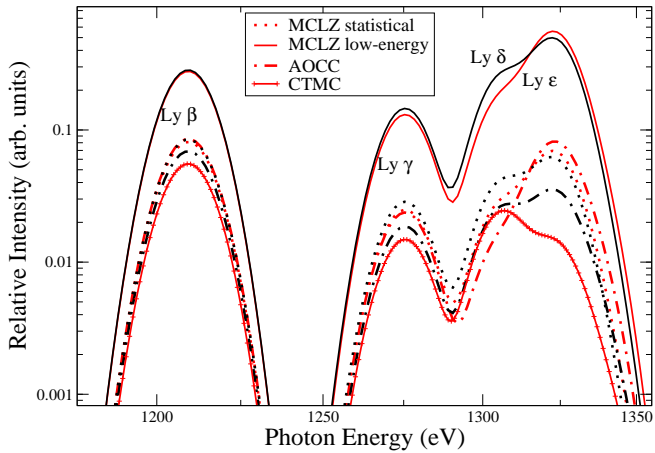}
\end{center}
\caption{ Ne$^{10+}$ + H X-ray spectra calculated using the AOCC and CTMC cross sections are compared to the MCLZ method at collisional energies of 1 keV/u 
(black lines) and 5 keV/u (red lines).}
\label{Fig3}
\end{figure}

\begin{figure}
\begin{center}
\includegraphics[scale=0.35, angle=270]{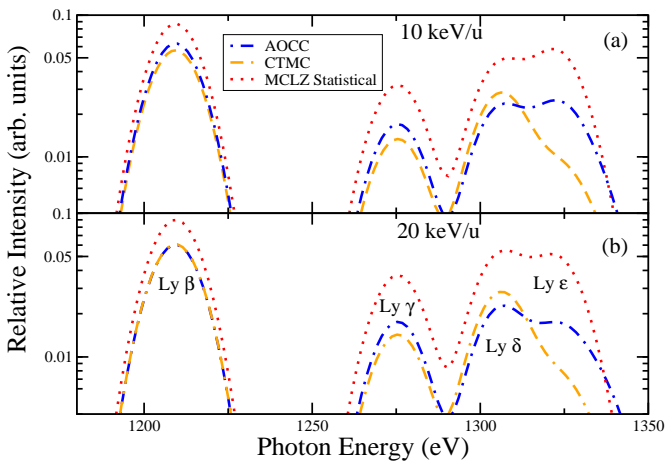}
\end{center}
\caption{The CTMC method is compared to the AOCC and MCLZ statistical $\ell$-distribution 
methods for the collision of  Ne$^{10+}$ with H at an energy of 10 keV/u (top panel) and 20 keV/u (bottom panel).}
\label{Fig4}
\end{figure}

Among the theoretical methods for the energy range and ion-atom systems considered here, at lower energies, 
the MCLZ low energy $\ell$-distribution is expected to describe the CX mechanism most accurately. Spectra are shown 
in Figure~\ref{Fig1} for Ne$^{10+}$ CX with H and He at collisional energies of 0.1, 0.5, and 1 keV/u using this method.\footnote{All line intensities
are normalized such that Ly$\alpha$ has an intensity of 1 and in most cases a Gaussian line-profile with a resolution of 10 eV full-width
half-maximum (FWHM) is adopted.} For Ne$^{10+}$ CX with H, 
as the collisional energy increases, the relative intensities of the Ly$\beta$ and higher lines show a slight increase. For Ne$^{10+}$ CX with He, the
trend is similar for the Ly$\gamma$ line, but reverses for the Ly$\delta$ line. The variation, however, is small. Further,
the Ly$\epsilon$ line is practically missing for \NeTen + He as the dominant $n$ capture channel for CX with He is typically one smaller than for CX with H.
Figure~\ref{Fig2} shows the similar spectra using the MCLZ statistical distribution. The trends are the same for CX with both H and He, however, for He, the higher Lyman lines have significantly smaller intensities giving a softer total X-ray spectrum. The statistical distribution is likely not valid at this low collision energy.
Figures~\ref{Fig3}-\ref{Fig4} show the CTMC, AOCC, and MCLZ low energy and statistical $\ell$-distribution spectra for 1-20 keV/u. The most notable
difference in comparison to the MCLZ results using the low-energy $\ell$-distribution is the decrease in the Ly$\epsilon$ line for all methods.

At higher energies (above $\sim$1-10 keV/u,) the statistical $\ell$-distribution\footnote{The statistical
   model assumes that the internal states are populated according to 
   their degeneracy which results from strong rotational mixing
   at small internuclear separation. Close internuclear distances
   become accessible for large kinetic energies.}   is expected to be more realistic in describing CX emission than 
the low energy $\ell$-distribution. Figure \ref{Fig7} shows the CX emission for \NeTen CX with He measured using simultaneous cold-target recoil ion
momentum spectroscopy and X-ray spectroscopy as described in \citet{ali10}
at 4.55 keV/u. The experimental single electron capture (SEC) CX emission is compared to spectra obtained with the CTMC \citep{ali05} and AOCC \citep{liu14} methods at the same energy, and the MCLZ low energy
and statistical $\ell$-distributions at 4.8 keV/u. Both MCLZ methods provide a better
fit to the experimental data than the CTMC or AOCC methods at this energy. The MCLZ statistical $\ell$-distribution results in slightly smaller Ly$\beta$ and higher lines, while
the MCLZ low-energy $\ell$-distribution results in slightly larger Ly$\beta$ and higher lines. At this resolution, the Ly$\beta$ and Ly$\gamma$ lines are
not clearly resolved in the theoretical spectra. 

In addition to SEC, observed and experimental X-ray emission spectra often contain contributions from
multiple electron capture (MEC), including 
double capture with subsequent autoionization (DCAI) and transfer ionization (TI). Emission from states populated by a combination of these 
reaction channels may be present
in astrophysical spectra. Figure \ref{Fig8} shows a comparison of DCAI to SEC for \NeTen CX with He using the CTMC method \citep{ali05}.
While TI has an insignificant contribution in the present ion-atom systems at low collision energy, DCAI 
is shown to give a contribution of 21$\%$ of the total spectrum for the CTMC method. 
However, Table 1 in \citet{ali05} shows that the experiment suggests a smaller 12$\%$ fraction of MEC, which agrees with the MCLZ method from that work. This
small fraction of MEC to the total CX emission suggests that ignoring MEC in \NeTen CX in an observed spectrum and assuming that SEC is dominant is justifiable
when considering He as the CX target, though exceptions for other ions are possible.

\subsection{Line Ratios}

The ratio of the intensity of the Ly$\beta$ and higher lines to the Ly$\alpha$ line vary for different processes and for different collision energies\footnote{Cross sections and line ratios are available upon request, but will be made available through the automated package Kronos \citep{mul16}, which contains a database of our preliminary H-like and He-like MCLZ CX cross sections.}.
These ratios can give insight into what
processes may be occurring within astrophysical environments, including, but not limited to, extracting the collisional energies
and ion abundance ratios.
Figures~\ref{Fig9}-\ref{Fig10} show the ratios of Ly$\beta$/Ly$\alpha$ calculated over different collisional energies
using the four different methods described in \S 2.1. 
Because the astrophysical environments of present interest contain  both neutral H and He, line ratios for the collisions of \NeTen +H (Fig.~\ref{Fig9}), 
He (Fig.~\ref{Fig10}), and for the assumed fraction of 90\% H  and 10\% He (Fig.~\ref{Fig10} for the MCLZ calculations) are shown.

\begin{figure}
\begin{center}
\includegraphics[scale=0.35, angle=270]{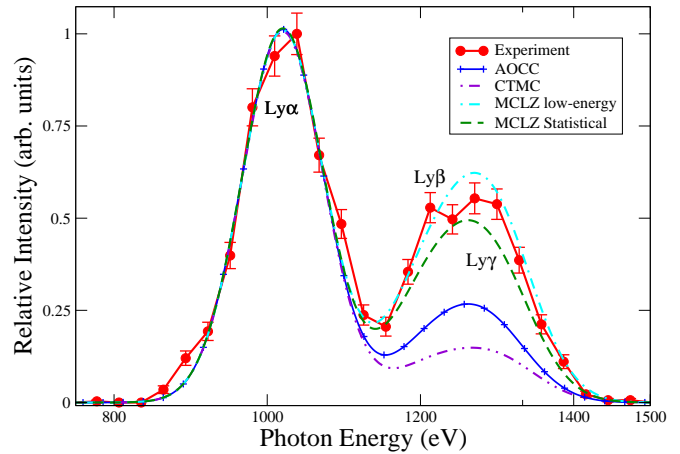}
\end{center}
\caption{Experimental single electron capture (SEC) data for Ne$^{10+}$ CX with He at $\sim$ 4.5 keV/u is compared to CTMC, AOCC \citep[][]{liu14},
and MCLZ low-energy and statistical distributions
for SEC. A resolution of 75 eV FWHM is used for the theoretical methods compared to a resolution of 68 eV for the experimental data
\citep{ali10}.
Note that the experimental spectra includes no multi-electron capture events. }
\label{Fig7}
\end{figure}

\begin{figure}
\begin{center}
\includegraphics[scale=0.35, angle=270]{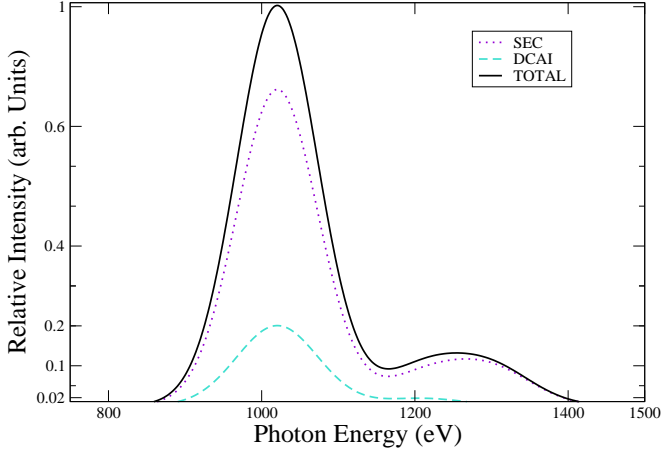}
\end{center}
\caption{Ne$^{10+}$ + He SEC is compared to DCAI and the summed total for the CTMC method with a 75 eV FWHM
resolution. SEC dominates the total with 79\% 
while DCAI contributes 21\% of the total. }
\label{Fig8}
\end{figure}

\begin{figure}
\begin{center}
\includegraphics[scale=0.35, angle=270]{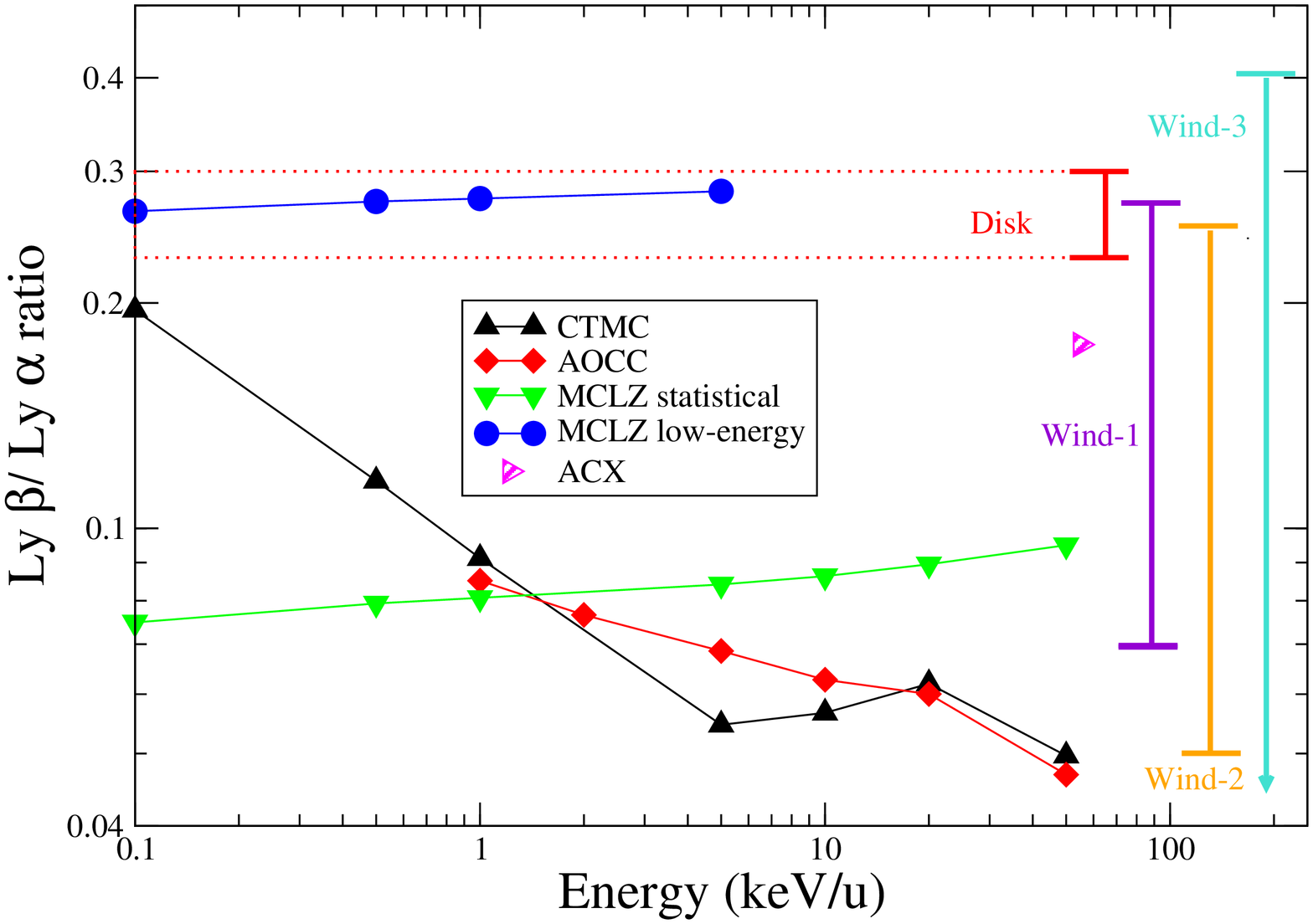}
\end{center}
\caption{Line ratios (Ly$\beta$/Ly$\alpha$) for the three methods shown in Figures~\ref{Fig2}-\ref{Fig4} as a function of 
collisional energy for charge exchange of Ne$^{10+}$ with H. The vertical lines represent the Disk, Wind-1, Wind-2, and Wind-3 regions
of M82 as described by \citet{kon11}. The results from the ACX model \citep{zha14} are also shown. Note that the results from
\citet{kon11} and \citet{zha14} have no ion kinetic energy dependence and their placement on the right side of the plot is arbitrary.}
\label{Fig9}
\end{figure}

\begin{figure}
\begin{center}
\includegraphics[scale=0.35, angle=270]{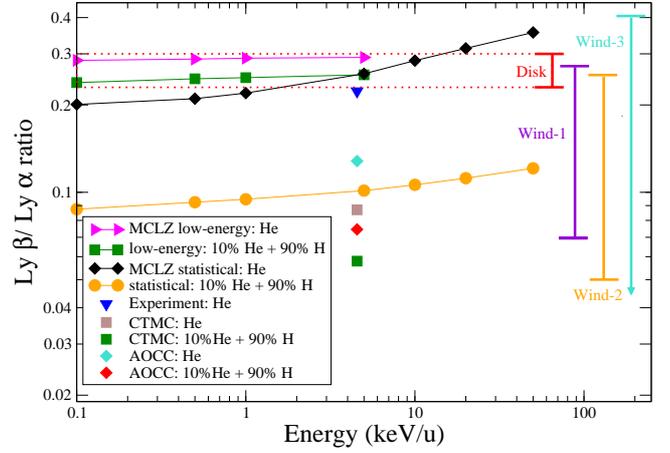}
\end{center}
\caption{Same as Figure~\ref{Fig9} but for the collision of Ne$^{10+}$ with He. Also shown is the addition Ne$^{10+}$ with 90\% H and 10\% He, which is
the assumed ratio of the He to H charge exchange contributions. CTMC, AOCC, and experimental Ne$^{10+}$ + He data at 4.55 keV/u (see Figure~\ref{Fig7})
are also compared.}
\label{Fig10}
\end{figure}

For Ne$^{10+}$ CX with H, the MCLZ statistical $\ell$-distribution and AOCC line ratios are almost equal at 1 keV/u, and are relatively
consistent between 1 and 10 keV/u. The CTMC method is very similar to the AOCC results for the range of 10-50 keV/u considered.
The MCLZ low energy $\ell$-distribution method, however, gives a line ratio over an order of magnitude larger than all other methods
for the entire energy range. 
The low energy $\ell$-distribution is expected to be more relevant at lower energies than the AOCC or CTMC methods, as suggested in Figure~\ref{Fig7}. 
However, extending the CTMC method to lower energies, it appears to approach the MCLZ low energy $\ell$-distribution for $E \la$ 0.1 keV/u. 
It can then be postulated that the transition from statistical to
low-energy $\ell$-distributions occurs between $\sim$0.1 and 1 keV/u making the MCLZ low energy $\ell$-distribution relevant at energies less than $\sim$0.5 keV/u. 


For Ne$^{10+}$ CX with helium, MCLZ cross
sections have been computed in the present work down to low energy ($\sim$100 eV/u)
and CTMC results at one impact energy \citep[4.5 keV/u,][]{ali10} are available from previous
work, though they could easily be extended to higher and lower energies. These results are shown in Figure~\ref{Fig10}
along with the experimental \citep{ali10} and the AOCC \citep{liu14} results at 4.55 keV/u.

\section{Application to M82}

In Figures~\ref{Fig9}-\ref{Fig10}, the Ly$\beta$ to Ly$\alpha$ ratios for Ne$^{10+}$ CX with H and He are overlaid with line ratio observations (vertical lines)
for M82 from {\it Suzaku} described by \citet{kon11}. The observations give the Ne$^{9+}$ Ly$\beta$ to Ly$\alpha$ ratio for four different regions in M82; 
the Disk, Wind-1, Wind-2,  and Wind-3. 
In their Figure 8, \citet{kon11} showed that the line ratios for the three Wind regions can be reproduced by a collisional ionization equilibrium (CIE) model, but
the Disk region cannot. The Ly$\beta$ to Ly$\alpha$ ratio from the CIE model is significantly smaller at $\sim$0.1 than the observed Disk line ratio ($\sim$0.25), indicating that the line ratio in the Disk might be primarily due to CX. 
Assuming that CX has a significant contribution to the spectrum of M82,
the collisional energies at which the Ly$\beta$ to Ly$\alpha$ line ratios cross the observed line ratio from \citet{kon11} gives 
a constraint on the ion kinetic energy in M82.\footnote{The Ne IX K$\beta$ line at 1074 eV may partially overlap the observed
Ne X Ly$\alpha$ line contaminating the line ratio. The K$\beta$ is expected, however, to account for less than $\sim$5\% of the Ne IX
emission and the removal of its contribution would only enhance the Ly$\beta$/Ly$\alpha$ ratio.}

Figures~\ref{Fig9} and \ref{Fig10} show that the MCLZ low energy $\ell$-distribution line ratios agree with the Disk region observation
by \citet{kon11} for H and He over a wide range of ion energies (0.1 to 5 keV/u), while the CTMC line ratio for H approaches the observational 
result near 0.1 keV/u.
The MCLZ low energy $\ell$-distribution is over an order of magnitude higher than all other methods
for the entire energy range in Figure~\ref{Fig9}, making it fall within the range of the Disk line ratio for M82. As postulated above,
the transition from statistical to a low-energy $\ell$-distribution occurs between 0.1 and 1 keV/u,
because the low energy $\ell$-distribution Ly$\beta$ to Ly$\alpha$ ratio
curve lies within the Disk region uncertainty, we can postulate that the ion kinetic energy in the disk is likely $\la$0.5 keV/u (300 km/s).

In the more unlikely case that all of the atomic H in the Disk region
has been ionized leaving only neutral He, Figure~\ref{Fig10} shows that again the
Ly$\beta$/Ly$\alpha$ ratio predicted by the MCLZ cross sections with
the low-energy $\ell$-distribution gives best agreement. The MCLZ statistical distribution for He predicts line ratios in agreement
with the {\it Suzaku} observation for ion velocities of $\sim$950-2000 km/s (1-10 keV/u).
Likewise, the measurements of \citet{ali10} performed
at 930 km/s are also consistent with the Disk observation, but these
high ion velocities likely cannot be produced in starburst outflows.
The CTMC and AOCC He results underestimate the observation by a factor of $\sim$3  at the energy for which data are available, namely, about 4.5 keV/u, 
which is considerably larger than the estimated 0.5 keV/u identified above as the likely maximum ion energy that is relevant to CX in the Disk region.

Considering CX from both H and He, but with abundance fractions of 90\% and
10\%, respectively, again the MCLZ low-energy distribution result agrees
best with the observations in the Disk region. For all other methods, He target
contributions increase the predicted Ly$\beta$/Ly$\alpha$ ratio, but only
to $\sim$0.1 at 4.55 keV/u (930 km/s). 

Using {\it XMM-Newton} observations with its Reflection Grating
Spectrometer (RGS), \citet{zha14} modeled the X-ray spectrum of M82
assuming a two component model consisting of hot single-temperature
thermal plasma emission (electron-impact excitation) and CX emission.
They found that the Ne X Ly$\alpha$ line intensity can be reproduced if 19\% of the
emission is due to CX. However, the RGS spectrum covers the entire disk and
outflow of M82 \citep[i.e., Disk, Wind-1, Wind-2, and Wind-3 regions of][]{kon11}. This is consistent with the suggestion of Konami
et al. that the Wind components can be modeled with thermal emission,
while the Disk is dominated by CX emission, the latter confirmed here.
Further, the CX contribution in \citet{zha14} was obtained with
the AtomDB Charge eXchange (ACX) model \citep{smi12} that adopted
the so-called separable $\ell$-distribution. The low-energy distribution
peaks at $\ell=1$, while the separable distribution peaks at
$\ell=2$ for Ne$^{10+}$ that results in the ACX Ly$\beta$/Ly$\alpha$ ratio of 0.176, shown in Figure~\ref{Fig9}.
This line ratio is somewhat smaller than the observed Disk ratio of $\sim$0.28 with {\it Suzaku}, but
consistent with total line ratio of 0.182 \citet{zha14} measured with the RGS. However, one must
remember the {\it XMM-Newton}/RGS measurement includes all regions
of the Disk and outflow.

To reproduce the observed Disk Ly$\beta$/Ly$\alpha$ ratio assuming
only CX emission, our modeling suggests that the relative ion velocity 
is $\la$300 km/s, independent of whether a 10\% He component is
included. While superwind
models of the M82 outflow typically adopt velocities greater than 1000
km/s, \citet{zha14} suggest that a reasonable approximation
is closer to $\sim$500 km/s. Finally, a further test of the contribution of
CX to the X-ray emission could be made by modeling the Ly$\delta$ and
Ly$\epsilon$ emission at 1308 and  1325 eV, respectively. Figure 2 suggests
that both line intensities (for H) are larger than or comparable to that of Ly$\beta$.
The Ly$\epsilon$ to Ly$\delta$ ratio could be used to constrain the
neutral He to H fraction given the negligible Ly$\epsilon$ emission for
Ne$^{10+}$ + He collisions. In fact, there is some hint of this feature
in the CX modeling of \citet{zha14}, though blending of the Mg XI K$\alpha$
forbidden line may contaminate the Ne lines. Higher resolution spectra, such as with \emph{Hitomi} will better resolve the emission lines and allow for better constraints on the contribution of Ne X CX.  A more thorough model, using for example, XSPEC \citep{arn96}, should be performed including H and He-like C, N, O, Ne, Mg, and Si and Fe XVII ions colliding with H and He targets, similar to the approach of \citet{zha14},
   but with reliable CX data. Such a comprehensive approach
   would help in determining the contribution of various mechanisms including, collisional
   excitation (CIE), radiative recombination, CX, or other
   processes such as active galactic nuclei (AGN) activity.

\section{Conclusions}

As shown here via our comparison of results of various conventional theoretical approaches, 
the MCLZ low-energy $\ell$-distribution is likely the better method for producing \NeTen CX line ratios with H and He below 5 keV/u. 
The statistical distribution may be used above 1 keV/u, but the AOCC method is preferred. As the AOCC and CTMC methods have
agreement between 10 and 50 keV/u, either method would work, but as the Ly$\epsilon$ lines (due to H) are similar for the MCLZ statistical and AOCC methods for this energy,
the AOCC approach is likely more accurate.

The current charge exchange modeling of the Ne X X-ray emission for
the Disk of the starburst galaxy M82 suggests that
CX may account for the observed Ly$\beta$/Ly$\alpha$ ratio,
a ratio that is not consistent with thermal emission due to electron-impact
excitation. Further, the emission may include contributions from CX
with neutral H and He suggesting a relative ion velocity of $\la$300
km/s. While cross sections obtained with the multichannel Landau-Zener method
using the so-called low-energy $\ell$-distribution model predicts X-ray
line ratios consistent with the observations, calculations using the theoretical method known as the 
molecular-orbital close-coupling approach, recognized to be among the most complete treatments of ion-atom collisions at low-impact
energies, but with significantly greater computational effort and complexity, should be made to confirm these results.

\section*{Acknowledgments}
The work of RSC, DL, DRS, and PCS was partially supported by NASA grants
NNX09AC46G and NNX13AF31G. RSC acknowledges support from NSF East Asian and Pacific Summer Institute and the
Chinese Academy of Sciences and thanks those at the IAPCM for hosting her at the beginning
of this project as well as Adam Foster, Randall Smith, and Shuinai Zhang for helpful discussions. We thank the referee for providing useful comments which improved the manuscript.

\label{lastpage}

\end{document}